\documentclass[preprint,showpacs,amsmath,amssymb,prd,nofootinbib]{revtex4-1}
\usepackage{epsf}
\usepackage{dcolumn}
\usepackage{amsmath}
\usepackage{amssymb}
\usepackage{color}
\usepackage[dvipsnames]{xcolor}
\usepackage{bm}
\everymath{\displaystyle}

\usepackage{graphicx}

\usepackage{cancel} 

\usepackage[breaklinks=true,colorlinks=true]{hyperref}

\begin{document}

\title{Quasi-uniform gravitational field of a disk revisited}

\author{Alexander J. Silenko\footnote{Email: alsilenko@mail.ru}}

\affiliation{Institute of Modern Physics, Chinese Academy of
Sciences, Lanzhou 730000, China,\\ Bogoliubov Laboratory of Theoretical Physics, Joint Institute for Nuclear Research,
Dubna 141980, Russia,\\
and Research Institute for Nuclear Problems, Belarusian State University, Minsk 220030, Belarus}

\author{Yury A. Tsalkou\footnote{Email: skytec@mail.ru}}

\affiliation{Research Institute for Nuclear Problems, Belarusian State University, Minsk 220030, Belarus}


\begin{abstract}
We calculate the quasi-uniform gravitational field of a disk in the weak-field approximation and demonstrate an inappropriateness of preceding results. The Riemann tensor of this field is determined. The nonexistence of the uniform gravitational field is proven without the use of the weak-field approximation. The previously found difference between equations of motion for the momentum and spin in the accelerated frame and in the quasi-uniform gravitational field also takes place for the disk. However, it does not violate the Einstein equivalence principle because of the nonexistence of the uniform gravitational field.
\end{abstract}

\maketitle


\section{Introduction}

The problem of a uniform gravitational field is rather important. This field has been used by Einstein \cite{Einstein,Einstein2,Einstein3} for the formulation of the equivalence principle. In addition, the uniform field is the simplest example of a gravitational field. Therefore, a detailed analysis of the problem of the uniform gravitational field is necessary for a right understanding of fundamentals of general relativity (GR).

In the present work, we give a review of previous studies of this problem. Contrary to most of these studies, we affirm and prove that the uniform gravitational field does not exist. We rigorously derive the metric of a quasi-uniform gravitational field of a disk in the weak-field approximation and show its substantial difference from formerly proposed metrics of the uniform gravitational field.
The results obtained confirm the conclusion previously made in Ref. \cite{PRDThomaspre} that criticisms based on different dynamics of a four-momentum and a spin of a test particle in a uniformly accelerated frame and in the Schwarzschild spacetime \cite{Synge,Fock,OhanianRuffini,Desloge,Gron,Moreau,VarjuRyder,Lee} do not violate the  Einstein equivalence principle (EEP). We discuss the issues related to the EEP.

The paper is organized as follows. The review of previous results on a uniform/quasi-uniform gravitational field is presented in Sec. \ref{reviews}.
In Sec.~\ref{quasiuniform}, we calculate the quasi-uniform gravitational field of a disk in the weak-field approximation. The Riemann tensor of this field is determined in Sec. \ref{Riemanntenso}. A nonexistence of the uniform gravitational field is proven in Sec. \ref{existence}.
The results obtained are discussed and summarized in Sec.~\ref{discussion}. 

We denote world and spatial indices by Greek and Latin letters
$\alpha,\mu,\nu,\ldots$ $=0,1,2,3$, $i,j,k,\ldots=1,2,3$, respectively. Tetrad
indices are denoted by Latin letters from the beginning of the
alphabet, $a,b,c,\ldots = 0,1,2,3$. Temporal and spatial tetrad indices are
distinguished by hats. The signature is $(+---)$.
The Ricci scalar curvature is defined by
$\mathcal{R}=g^{\mu\nu}R_{\mu\nu}=g^{\mu\nu}R^\alpha_{~\mu\alpha\nu}$, where
$R^\alpha_{~\mu\beta\nu}=\partial_\beta\Gamma^\alpha_{~\mu\nu}-\ldots$
is the Riemann curvature tensor.
Commas and semicolons before indices denote partial and covariant derivatives, respectively. We suppose that the coordinates $x^\mu$ are always components of a contravariant vector. We use the system of units with
$\hbar=1,~c=1$. These constants are explicitly displayed
in some formulas when their introduction clarifies a description of physical phenomena.

\section{Review of previous results}\label{reviews}

The metric of the uniform/quasi-uniform gravitational field is an important problem of GR.
Probably, the first attempt to determine this metric has been made by Kottler \cite{Kottler}. This attempt was based on the assumption of the full similarity between the metrics of the uniform gravitational field and the uniformly accelerated frame  (the Rindler metric). The former metric has been obtained from the latter one \begin{equation} g_{\mu\nu}={\rm diag}\,\biggl(\Bigl[1+\frac{\bm a\cdot\bm r}{c^2}\Bigr]^2,\,-1,\,-1,\,-1\biggr)
\label{accel} \end{equation}
 by the replacement $\bm a\rightarrow -\bm g$. The same assumption has been made by M{\o}ller \cite{Moller}. Thus, the Kottler-M{\o}ller metric has the form \begin{equation} g_{\mu\nu}={\rm diag}\,\biggl(\Bigl[1-\frac{\bm g\cdot\bm r}{c^2}\Bigr]^2,\,-1,\,-1,\,-1\biggr).
\label{Kottler} \end{equation} An inertial force acting on an observer in the uniformly accelerated frame is antiparallel to the frame acceleration. This explains the different signs in Eqs. (\ref{accel}) and (\ref{Kottler}).

However, this result was not generally accepted to be satisfactory. Trivial spatial components in the metric (\ref{Kottler}) cannot be explained on the basis of the Schwarzschild metric. Next investigations have resulted in the Kottler-Whittaker metric \cite{Kottler2,Whittaker} which is given by
\begin{equation} g_{\mu\nu}={\rm diag}\,\biggl(1+\frac{2gx}{c^2},\,-\Bigl[1+\frac{2gx}{c^2}\Bigr]^{-1},\,-1,\,-1\biggr).
\label{Kottler2} \end{equation} The same result has been obtained by Rohrlich \cite{Rohrlich} among admissible metrics.
The metric (\ref{Kottler2}) has been rediscovered by Krige \cite{Krige}.

The both metrics, (\ref{Kottler}) and (\ref{Kottler2}), were much discussed (see Refs. \cite{Rohrlich,Jones,Munoz,Apple} and references therein).
We can also mention the important property proven in Refs. \cite{Rohrlich,Tilbrook}. A metric of the form
$$ds^2 =\lambda^2(x)c^2dt^2 -\sigma^2(x)dx^2-dy^2-dz^2$$
is flat only if
\begin{equation} \sigma(x)=\frac{1}{\mathfrak{g}}\frac{d\lambda(x)}{dx},
\label{Rohrlich} \end{equation}
where $\mathfrak{g}\neq 0$ is an integration constant. We obtain the Kottler-M{\o}ller metric when $\lambda(x)=1+gx~(\mathfrak{g}=g)$ and the Kottler-Whittaker one when
$\lambda(x)=\sqrt{1+ gx }  ~ (\mathfrak{g}=g/2)$.

As a result, the metrics (\ref{Kottler}) and (\ref{Kottler2}) describe flat spaces. However, this property of the Kottler-M{\o}ller and Kottler-Whittaker metrics is controversial because any gravitational sources create curved spacetimes.

We should also mention the domain
wall metric first given by Taub \cite{Taub,Taub_51} and considered also by Vilenkin \cite{Vilenkin} and other authors:
\begin{equation} ds^2 =(1+K|z|)^{-1/2}(c^2dt^2 -dz^2)-(1+K|z|)(dx^2+dy^2),
\label{Taubm} \end{equation} where $K=const$. General properties of static plane-symmetric spacetimes have been investigated in Refs. \cite{Jones,Munoz,EKasner,Das,Amundsen,DaSilva}.
It should be noted that the solution \eqref{Taubm} describes a vacuum domain wall and there is a singularity at $|z|=-1/K,~K<0$.

We mention a strong impact of Refs. \cite{Taub,Taub_51,Vilenkin} on the contemporary cosmology. However, such papers do not directly relate to the problem under consideration.
As a rule, the presented analysis is restricted to
uniform plane-symmetric field configurations which metrics depend on the \emph{one} spatial coordinate only \cite{Taub,Taub_51,Vilenkin,EKasner}. 
A careful analysis of the problem needs, taking into account a dependence of the metric on \emph{three} spatial coordinates (see Secs. \ref{quasiuniform},\ref{Riemanntenso}).

Besides the Taub static solution, Davis and Ray have considered the vacuum homogeneous time-dependent plane symmetric solution \cite{Davis_Ray} which relates to the special case of the Kazner metric. It has been shown that there is no natural way to join a static and nonstatic solutions at the metric singularity as it is done for the Schwarzschild solution inside and outside the horizon.

The important general analysis of canonical forms for axial symmetric spacetimes has been carried out in  Refs. \cite{Carlson_Safko2,CARLSONSAFKO,Carlson_Safko}. However, this analysis does not give information about the special form of the metric tensor in the case under consideration.

The conception of the uniform gravitational field has been used in Einstein's formulation of the equivalence principle \cite{Einstein}. The validity of the EEP has been discussed many times. The situation is more controversial in the
relativistic domain. It is pointed out in most books that the EEP is applicable only locally \cite{Treder,MTW,Schutz,Stephani,RindlerRel,Carroll}, but some authors disclaim its validity even locally \cite{Synge,Fock,OhanianRuffini}. There is not a full agreement between research articles as well.
It has been claimed, in particular, in Refs. \cite{Desloge,Gron} that the uniformly accelerated frame and a frame at rest in the uniform gravitational field are nonequivalent. A difference between a deflection of light in an accelerated frame and a gravitational field has been found in Ref. \cite{Moreau}. In Ref. \cite{Rohrlich}, the EEP has been supported with serious disclaimers
\footnote{
The two basic assertions,
(A) and (B), regarded in Ref. \cite{Rohrlich} as ``the principle of equivalence'' are obviously right. 
}.
Some other works \cite{Munoz,Lo,Bernstein} advocate the EEP. Many works relate only to the weak equivalence principle (see, e.g., Ref. \cite{Nobili}). Main criticisms of the EEP have been collected and have been considered in Ref. \cite{Lo}.

The problem of the uniform gravitational field and EEP can be treated in a wider context.
For instance, it has been shown in Ref. \cite{Apple} (where non-inertial effects have been induced by \emph{torsion} in terms of \emph{teleparallelism}) how the 
uniform gravitational field can be constructed from spaces with a vanishing curvature.

Nowadays, the validity of the equivalence principle for spin effects has been much discussed. It has been claimed in Refs. \cite{VarjuRyder,Lee} that the EEP is violated due to a difference between the angular velocities of the spin precession in
the uniformly accelerated frame and in the Schwarzschild metric. In the nonrelativistic limit ($v\ll c$), the corresponding quantities are equal to \cite{Lee,MashhoonObukhov}
\begin{equation}\begin{array}{c}
\bm\Omega^{(a)}=-\frac{\bm a\times \bm v}{2c^2}, ~~~ \bm\Omega^{(i)}=\frac{3\bm g\times \bm v}{2c^2},
\end{array}\label{equnLee}\end{equation} where $\bm v$ is the velocity of the spinning particle.
It has been claimed in Refs. \cite{VarjuRyder,Lee} that the EEP is violated owing to the threefold difference between the two angular velocities on condition\footnote{Equation (\ref{equnLee}) follows from the results obtained in Ref. \cite{VarjuRyder} after correcting a computational error, see Ref. \cite{MashhoonObukhov}.
} that $\bm a=-\bm g$. 
In Ref. \cite{PRD}, relativistic formulas for $\bm\Omega^{(a)}$ and $\bm\Omega^{(i)}$ have been obtained. Relativistic equations of the particle motion in the uniformly accelerated frame and in the Schwarzschild metric in the isotropic coordinates which have also been derived in Ref. \cite{PRD} differ as well. We can mention the full agreement between the classical and quantum-mechanical results \cite{PRD,Warszawa,OST,OST2,OST3,OST4,OST5,OST6,OST7,OST8}.

Thus, the equations of motion for the four-velocity (four-momentum) and for the spin obtained in the weak-field approximation in the uniformly accelerated frame and in the Schwarzschild field really differs. Nevertheless, it has been certified in Ref. \cite{PRDThomaspre} that their differences do not violate the EEP.  In Einstein's papers \cite{Einstein},
the equivalence principle has been formulated only relative to constant \emph{uniform} gravitational fields. The Schwarzschild field (as well as other real gravitational fields) is \emph{nonuniform}. The spatial inhomogeneity significantly influences the form of the equations of motion. The terms different for the Schwarzschild field and the uniformly accelerated frame ($\bm g=-\bm a$) are of the same order of magnitude as the corresponding terms different for the Schwarzschild field in the Cartesian and isotropic coordinates \cite{PRDThomaspre}.  Therefore, the difference between the equations of motion in the Schwarzschild field and the uniformly accelerated frame does not violate the EEP. This problem will be considered in detail in Sec. \ref{accelerationgravity}.

\section{Uniform and quasi-uniform fields: electrodynamics versus gravity}\label{quasiuniform}


The metrics of the uniform gravitational field presented in Refs. \cite{Kottler,Moller,Kottler2,Whittaker,Krige,Rohrlich,Jones,Munoz} are not obtained as a result of rigorous derivations of a metric defined by a known distribution of gravitational sources.
To fulfil such a derivation, it is instructive to compare uniform and quasi-uniform fields in electrodynamics and gravity.

\subsection{Fields of a charged disk and a charged infinite plane}

A uniform electric field can be created by electric charges distributed over an infinite plane. 
An electric field of a charged disk with the radius $R$ is quasi-uniform near its surface $z\ll R$. We suppose the surface charge density $\sigma$ to be constant. We apply the cylindrical coordinates $\rho,\varphi,z$ and the approach based on a summation of fields of separate charges. A similar approach based on the weak-field approximation can be used in gravity. In both electrodynamics and gravity, one needs to calculate dynamics of a test particle in external fields. The Hamiltonian of the test particle with the charge $e$ in an electrostatic field is given by
\begin{equation}
H=\sqrt{m^2+\bm p^2}+e\Phi,
\label{Hame} \end{equation} where $\bm p=m\gamma\bm v$ is the particle momentum, $\gamma$ is the Lorentz factor, and $\Phi$ is the scalar potential.
The equation of motion reads
\begin{equation}
\frac{d\bm p}{dt}=-\nabla H=e\bm E,
\label{eqmot} \end{equation}
where $\bm E=-\nabla\Phi$ is the electric field strength. The similar equation defines the particle acceleration $\bm w$:
\begin{equation}
\bm w=\frac{d\bm v}{dt}=\frac{e}{m\gamma}\left[\bm E-\bm\beta(\bm\beta\cdot\bm E)\right],\quad \bm\beta=\frac{\bm v}{c}.
\label{eqacc} \end{equation}

Equations (\ref{eqmot}) and (\ref{eqacc}) describing dynamics of the charged test particle contain derivatives of the scalar potential and do not contain this potential. The quantity $\Phi=\sum_i{\Phi^{(i)}}$ is equal to the sum of Coulomb potentials defined by the charges $q^{(i)}$ on the plane. These potentials are given by \begin{equation}
\Phi^{(i)}=\frac{q^{(i)}}{|\bm r-\bm r^{(i)}|},\quad \bm r=(x,y,z),\quad \bm r^{(i)}=(x^{(i)},y^{(i)},0).
\label{Coulomb} \end{equation}
When $x=y=0$, this potential depends on the only variable $z$ and the $z$ component of the field strength reads
\begin{equation}
E_z=-\sum_i{\frac{d\Phi^{(i)}}{dz}}=\sum_i{\frac{q^{(i)}z}{\sqrt{{x^{(i)}}^2+{y^{(i)}}^2+z^2}}}.
\label{CoulombE} \end{equation}

We can express the charge of a small area in the form $dq=\sigma\rho\, d\rho\, d\varphi$. For the charged disk, the integration over these coordinates leads to the well-known relations
\begin{equation}\begin{array}{c}
\Phi=2\pi\sigma\left(\sqrt{R^2+z^2}-|z|\right),\\
 E_z=2\pi\sigma\left(\mathrm{sign}(z)-\frac{z}{\sqrt{R^2+z^2}}\right),
\end{array}\label{Thomprespr}\end{equation}
where $\mathrm{sign}(z)$ is the signum function.

For the infinite plane,
\begin{equation}
\Phi=-2\pi\sigma |z|+\Phi_0,\qquad  E_z=2\pi\sigma\, \mathrm{sign}(z).
\label{Coulombpot} \end{equation}
If the potential energy of a free charged particle (being beyond any fields) is arranged to be zero, $\Phi_0$ is infinite. In electrodynamics, this problem can be solved with a redefinition of the potential energy. In particular, it can be supposed to be equal to zero at $z=0$. However, the similar problem in gravity is much more difficult.

\subsection{Field of a gravitating disk}
\label{disk_field}

The problem of a gravitational field of a uniform disk in GR is not so straightforward as in the Newtonian gravity.
Recall that the general form of an axially symmetric static solution of the vacuum Einstein equation
has a Weyl form (see, for instance, Ref. \cite{Synge}):
\begin{equation}\label{dsaxial_0}
	ds^2 = e^{2\psi}dt^2 - e^{2(\sigma-\psi)}(d\rho^2+dz^2)-\rho^2 e^{-2\psi} d\varphi^2,
\end{equation}
where $\{t,\rho, z,\varphi\}$ are so-called canonical (quasicylindrical) coordinates and $\psi$ and $\sigma$ are functions of $\rho,z$ only.
For this metric, vacuum field equations reduce to the Laplace equation for $\psi$:
\begin{equation}\label{dsaxial_1}
	\nabla^2 \psi = \psi_{,\rho\rho}+\frac{\psi_{,\rho}}{\rho}+\psi_{,zz}=0,
\end{equation}
and
\begin{equation}\label{dsaxial_2}
 \left\{\begin{matrix}
	\sigma_{,\rho} =& \rho(\psi^2_{,\rho}-\psi^2_{,z}), \\
	\sigma_{,z} =& 2 \rho \, \psi_{,\rho} \psi_{,z}.
\end{matrix}\right.
\end{equation}

Equations \eqref{dsaxial_0}--\eqref{dsaxial_2} seem to be simple. The problem is, however, to establish a connection between the source of the field and the canonical potential $\psi$ \cite{bonnor1992physical,voorhees1970static,Chou1931}.
For instance, the Schwarzschild solution in the form \eqref{dsaxial_0} has the potential $\psi$ corresponding (in canonical coordinates) to the Newtonian potential of a finite rod with the length equal to the gravitational radius of the central body \cite{griffiths2009exact}.

In this picture, a uniform disk represents some singular solution which representation needs an infinite number of terms. Thus, it is more convenient to consider non-uniform disk configuration \cite{Morgn,Morgn2,Azuma,Bicak1993,letelier1987exact}.

To avoid these problems, we restrict ourselves to the consideration of the GR weak-field approximation, where some similarity with electrodynamics appears.


In this approximation, the metric tensor takes the form
%
%
%
%
\begin{equation}\begin{array}{c}
g_{\mu\nu}=\eta_{\mu\nu}+h_{\mu\nu}, \qquad h_{\mu\nu}=\sum_k{h_{\mu\nu}^{(k)}}.
\end{array}\label{metrtensor}\end{equation}
Since $|h_{\mu\nu}|\ll1$, the Einstein equations become linear and the fields
$h_{\mu\nu}^{(k)}$ 
satisfy the principle of superposition (see Ref. \cite{LL2}).

The Hamiltonian of a test particle is given by \cite{Cogn}
\begin{equation}
H = \left(\frac{m^2 - G^{ij} p_ip_j} {g^{00}}\right)^{1/2} -
\frac{g^{0i}p_i}{{g}^{00}},\qquad G^{ij}=g^{ij} - \frac{g^{0i}g^{0j}}{{g}^{00}}.\label{clCog}
\end{equation}

The equations of motion in GR read
\begin{equation}
\frac{dp_\mu}{dt}=\frac{\partial H}{\partial x^\mu}, \label{clCop}
\end{equation}
\begin{equation}
\begin{array}{c}
\frac{du_{\mu}}{ds}=\frac12g_{\nu\lambda,\mu}u^\nu u^\lambda,
\end{array} \label{Lorcova} \end{equation} and
\begin{equation}
\frac{du^{\mu}}{ds}=-\left\{^{\,\mu}_{\nu\lambda}\right\}u^\nu u^\lambda,
 \label{LorChrc} \end{equation}
where
\begin{equation}
\begin{array}{c}
\left\{^{\,\mu}_{\nu\lambda}\right\}=\frac12g^{\mu\rho}\left(g_{\rho\nu,\lambda}+g_{\rho\lambda,\nu}-g_{\nu\lambda,\rho}\right)
\end{array} \label{Chrcgen} \end{equation} are the Christoffel symbols.
Equations (\ref{clCog}), (\ref{Lorcova}), and  (\ref{LorChrc}) show that the particle motion is defined by the derivatives of the metric tensor and is contributed by \emph{all} components of this tensor.

To calculate the metric tensor, we need to use the known equation 
for the metric of a weak Schwarzschild source in the Cartesian coordinates $(x,y,z)$ numerated as $(1,2,3)$: 
\begin{equation}\begin{array}{c}
 h_{00}^{(k)}=2\phi^{(k)}, \quad \phi^{(k)}=
 -{\frac {GM^{(k)}} {c^2|\bm r-\bm r^{(k)}|}}, \qquad h_{0i}^{(k)}=0, \\
 h_{ij}^{(k)}=\frac{2\phi^{(k)} \left(\bm r-\bm r^{(k)}\right)^{i}\left(\bm r-\bm r^{(k)}\right)^{j}}{|\bm r-\bm r^{(k)}|^2},
\end{array}\label{wip}\end{equation} 
%
%
where $M^{(k)}$ is the mass of the $k$-th source and $\phi^{(k)}$ is its  
Newtonian potential in $c^2$ units. Evidently, 
$h_{11}^{(k)}+h_{22}^{(k)}+h_{33}^{(k)}=h_{00}^{(k)}$.

We consider the static metric and the gravitational field on the axis of the disk ($x=y=0$). In this case,  $g_{\mu\nu,0}=0$,
but first and second derivatives of the metric tensor with respect to spatial coordinates can be nonzero.
The mass of a small area can be expressed in the form $dM=\mu\rho\, d\rho\, d\varphi$, where $\mu$ is the surface mass density which is supposed to be constant. Next derivations are similar to those in electrodynamics. In particular, $g_{00}$ is equal to [cf. Eq. (\ref{Thomprespr})]
\begin{equation}\begin{array}{c}
 g_{00}=1 + h_{00}=1 +2\phi,\qquad \phi= -\frac {2\pi G\mu} {c^2}\left(\sqrt{R^2+z^2}-|z|\right).
\end{array}\label{wipgrav}\end{equation}

The other nonzero metric components are $g_{11},g_{22}$, and $g_{33}$. They are defined by
\begin{equation}\begin{array}{c}
 h_{11}=h_{22}=-\frac {2\pi G\mu} {c^2}\cdot\frac{\left(\sqrt{R^2+z^2}-|z|\right)^2}{\sqrt{R^2+z^2}},\\
 h_{33}=-\frac {4\pi G\mu} {c^2}\cdot \left(|z|-\frac{z^{2}}{\sqrt{R^2+z^2}}\right).
\end{array}\label{wipnext}\end{equation} The spatial components $g_{ij}=0$ when $i\neq j$ and $x=y=0$. It can be checked that $h_{11}+h_{22}+h_{33}=h_{00}$.



Evidently, $h_{11}=h_{22}$ due to the cylindrical symmetry of the disk. But it is easy to show that the number of the independent metric components is even smaller:
\begin{equation}\label{g33depend}
h_{33}=h_{00}-2h_{11},\qquad
g_{33}=g_{00}-2g_{11}-4.
\end{equation}
Thus, $g_{\mu\nu}$ contains 
only two linear independent functions of $z$. 
Since we consider the metric near the symmetry axis, the tensor $g_{\mu\nu}$ does not \emph{explicitly} depend on horizontal coordinates. However, derivatives of the metric tensor with respect to the horizontal coordinates should be taken into account.

%
%


Note that there are other types of the canonical form of axial gravitational fields \cite{CARLSONSAFKO}, which relate to the Weyl form. A difference between them is mostly a matter of definition.

Here we emphasize some other properties of the metric \eqref{wipgrav}, \eqref{wipnext}.
First, 
it obviously retains plane symmetry and recovers the Schwarzschild metric in the far-distance limit. This limit contains total mass of the disk $M=\mu\, \pi R^2$.

Second, the gravitational potential $\phi$ from Eq. \eqref{wipgrav} coincides with that for a uniform disk in the exact Newtonian solution \cite{lass1983gravitational} at $\rho = 0$.	




%
%
%
%


Since the weak-field approximation is used,
$g^{00}=1-2\phi,~g^{11}=-1-h_{11},~g^{22}=-1-h_{22},~g^{33}=-1-h_{33}$,
and $g^{\mu\nu}=0$ when $\mu\neq\nu$. To specify the general
equations of motion (\ref{clCog}) and (\ref{Lorcova}), we need to
calculate the derivatives of the metric tensor components:
\begin{equation}\begin{array}{c} g_{00,3}=\frac {4\pi G\mu} {c^2}\left(\mathrm{sign}(z)-\frac{z}{\sqrt{R^2+z^2}}\right), \\
g_{11,3}=g_{22,3} =\frac {2\pi G\mu} {c^2}\left[2\,\mathrm{sign}(z)-\frac{3z}{\sqrt{R^2+z^2}} \right.\\ \left.+\frac{z^3}{(R^2+z^2)^{3/2}}\right],\\
g_{33,3}=-\frac {4\pi G\mu} {c^2}\left[\mathrm{sign}(z)-\frac{2z}{\sqrt{R^2+z^2}}+\frac{z^3}{(R^2+z^2)^{3/2}}\right].
\end{array}\label{wipnder}\end{equation}

In order to obtain other derivatives with respect to the spatial coordinates, one needs 
to take
a step back and to
find the respective derivatives from \eqref{wip}. After this, one needs to integrate again on the whole disk like in Eqs. \eqref{wipgrav}, \eqref{wipnext}. For instance,
\begin{equation}\label{gnozero0}
g_{13,2}= \sum_k~\frac{6  G M^{(k)} (x-x^{(k)})~(y-y^{(k)})~z }{c^2\left[(x-x^{(k)})^2+(y-y^{(k)})^2+z^2\right]^{5/2}}.
\end{equation}
It is easy
to show that remaining nonvanishing derivatives read
\begin{equation}\label{gnozero}
g_{13,1}=g_{31,1}=g_{23,2}=g_{32,2}= -\frac{2 \pi G z \mu R^2 }{c^2 (R^2+z^2)^{3/2}}.
\end{equation}
The metric derivatives have only two linearly independent components. For instance, in terms of $g_{00,3}$  and $g_{31,1}$ one obtains the following relations:
\begin{equation}\begin{array}{c}\label{gmnDerivDepend}
g_{11,3}=g_{22,3}=g_{00,3}+g_{31,1},\\
g_{33,3}=-g_{11,3}-g_{31,1}=-g_{00,3}-2g_{31,1}.
\end{array}
\end{equation}	
Obviously, this is the direct consequence of the metric symmetry (see Eq.  \eqref{g33depend}). It is well-known that the particle motion is defined by the Christoffel symbols proportional to first derivatives of the metric tensor. Therefore, a dependence of the metric tensor on the horizontal coordinates significantly influences the particle motion.

As it is expected for plane symmetry of the disk, all the metric components \eqref{wipgrav} and \eqref{wipnext} are even and all the metric derivatives \eqref{wipnder} and \eqref{gnozero} are odd with respect to $z$.

The Newtonian acceleration is equal to
\begin{equation}\begin{array}{c} \bm g=-\frac{c^2}{2}g_{00,3}\bm e_z=-2\pi G\mu\left(\mathrm{sign}(z)-\frac{z}{\sqrt{R^2+z^2}}\right)\bm e_z.
\end{array}\label{Newtonian}\end{equation}

The Christoffel symbols which are nonzero on the symmetry axis of the disk are given by $\left(\left\{^{\,\mu}_{\nu\lambda}\right\}=\left\{^{\,\mu}_{\lambda\nu}\right\}\right)$
\begin{equation}\begin{array}{c}
\left\{^{\,0}_{30}\right\}=\frac12g_{00,3},\quad \left\{^{\,1}_{31}\right\}=-\frac12g_{11,3},\quad \left\{^{\,2}_{32}\right\}=-\frac12g_{22,3},\\ \left\{^{\,3}_{00}\right\}=\frac12g_{00,3},\quad \left\{^{\,3}_{11}\right\}=\frac12g_{11,3}
-g_{31,1},
\quad \left\{^{\,3}_{22}\right\}=\frac12g_{22,3}
	-g_{32,2},
\\
\left\{^{\,3}_{33}\right\}=-\frac12g_{33,3}.
\end{array} \label{Chrcgenn} \end{equation}

To check how the solution \eqref{wipgrav}, \eqref{wipnext} obeys the junction condition for singular hypersurfaces  	\cite{Israel1966}, let us consider the surface energy-momentum $3\times3$-tensor $S_{ik}$ induced on the disk. Recall that  $S_{ik}$ is expressed via the extrinsic curvature $K_{ik}$ as follows \cite{MTW}:
	\begin{equation}\label{Sik}
	\kappa \epsilon \, S_{ik}=\left([K_{ik}] -\tilde{h}_{ik} [K^i_{\;i}] \right),
	\end{equation}
	where $\kappa = 8 \pi G/c^4$, $\tilde{h}_{ik} = g_{ik}-\epsilon \, n_i n_k$ is a projection of the metric to the surface, $\epsilon=-1$ for the spacelike unit normal to the disk surface, $n_\mu$. The value in the square brackets designates its jump across  the disk:  $[K]\equiv K^+-K^-$. In the current setup, $z$ direction is orthogonal to the disk surface. Thus, coordinate system is normal Gaussian and we get (see Refs. \cite{MTW,GronHervik})
	\begin{equation}\label{}
	K_{ik}^\pm
	=\epsilon \lim_{z\rightarrow 0\pm 0}
	\left( n_\alpha \left\{^{\,\alpha}_{ik}\right\}\right)
	=\epsilon \lim_{z\rightarrow 0\pm 0}
	\left( n_3 \left\{^{\,3}_{ik}\right\}\right)
	=\frac{\epsilon}{2} \lim_{z\rightarrow 0\pm 0} \frac{\partial g_{ik}}{\partial z},
	\end{equation}
	where Eqs. \eqref{Chrcgenn} and vanishing of derivatives of the nondiagonal metric component at $z \rightarrow 0$ are taken into account [see Eq. \eqref{gnozero}].
	With the use of Eqs. \eqref{wipnder}, we can check that
	\begin{equation}\label{Sdisk}
	S_{ik}=
	\begin{pmatrix}
	\mu c^2 & 0 & 0 \\
	0 & 0 & 0 \\
	0  & 0  & 0   \\
	\end{pmatrix}
	\end{equation} for r.h.s. of Eq. \eqref{Sik} in the leading order on $G/c^2$.
From the point of view of the equation of state (EoS), we deal, as expected, with a pressureless dense dust with the surface density 
$S_{00}/c^2=\mu >0$. Thus, the metric \eqref{wipgrav}, \eqref{wipnext} satisfies the junction condition of geometrical embeddings.
	%

Additional circumstance following from Eq. \eqref{Sdisk} is that energy-momentum tensor of our disk has usual time-like structure, which differs from the solutions based on the Taub and Kazner metrics \cite{Taub,Taub_51,EKasner}, where the source is
thin null pressureless fluid shocks \cite{Carlson_Safko}.	

It is easy to introduce the cylindrical coordinates $\rho,\varphi,z$, because
\begin{equation}
g_{11}dx^2+g_{22}dy^2=g_{11}(dx^2+dy^2)=g_{11}(d\rho^2+\rho^2d\varphi^2). \label{Chrcglm} \end{equation}

We can note that the metric of the static disk in the weak-field approximation defined by Eqs. (\ref{wipgrav}),  (\ref{wipnext}), and (\ref{Chrcglm}) differs from the metric \eqref{dsaxial_0}
which is generally used for a description of axially symmetric spacetimes (see Refs. \cite{Synge,Morgn,Morgn2,Azuma} and references therein).
The quasicylindrical coordinates applied in Eq.
\eqref{dsaxial_0} 
substantially differ from the genuine cylindrical coordinates.
Nevertheless, it is possible to obtain a solution for a weak disk field in all space in canonical quasicylindrical coordinates and relate it to the solution \eqref{wipgrav}, \eqref{wipnext}. This problem is 
discussed in the Appendix. 	


\subsection{Field near a disk surface}

To compare our formulas rigorously derived in the weak-field approximation with the results obtained in Refs.
\cite{Kottler, Moller,Kottler2,Whittaker,Krige,Rohrlich,Jones,Munoz} for an infinite gravitating plane, we need to consider the field near a disk surface ($|z|\ll R$). In this case, the metric components and their first derivatives are given by
\begin{equation}\begin{array}{c}
 g_{00}=1-\frac {4\pi G\mu} {c^2}\left(R-|z|\right),\\  g_{11}=g_{22}=-1
-\frac {2\pi G\mu} {c^2}\left(R-2|z|\right),\\
g_{33}=-1-\frac {4\pi G\mu} {c^2}|z|,
\end{array}\label{wipgnear}\end{equation}
\begin{equation}\begin{array}{c}
 g_{00,3}=
 g_{11,3}=g_{22,3}=-g_{33,3}=
\frac {4\pi G\mu} {c^2}\,\mathrm{sign}(z).
\end{array}\label{wipdnear}\end{equation}

Evidently, neither the metric tensor nor its first derivatives agree with the previous investigations
\cite{Kottler, Moller,Kottler2,Whittaker,Krige,Rohrlich,Jones,Munoz,Apple} where the metric of a uniform gravitational field was searched. The discrepancy takes place even in the limit of $R\rightarrow\infty$.
To solve some problems related to the EEP, we need also to calculate the Riemann tensor.

\section{Riemann tensor}\label{Riemanntenso}

Since the field uniformity is approximate but is not exact, the
spacetime is not flat. It is important to determine the Riemannian
curvature. In the weak-field approximation, the definition of the
Riemann tensor reduces to
\begin{equation}
R^{\mu}_{~\nu\lambda\rho} = \left\{_{\nu\rho}^{\,\mu}\right\}_{,\lambda}- \left\{_{\nu\lambda}^{\,\mu}\right\}_{,\rho}.
\label{Riemann}
\end{equation}
It can be shown that in this approximation
\begin{equation}
R_{\mu\nu\lambda\rho} =\frac12 \left(g_{\mu\rho,\lambda,\nu}-g_{\mu\lambda,\rho,\nu}+g_{\nu\lambda,\rho,\mu}-g_{\nu\rho,\lambda,\mu}\right).
\label{Riemannweakfieldapp}
\end{equation}
Equations (\ref{Chrcgen}), (\ref{Riemann}), and (\ref{Riemannweakfieldapp}) allow us to obtain some nonzero components of the Riemann tensor:
\begin{equation} R_{0330} =- R_{3030} =\frac12g_{00,3,3}=-\frac {2\pi G\mu R^2} {c^2(R^2+z^2)^{3/2}}.
\label{Riemann30}
\end{equation}
To calculate all nonzero components, we need to determine
derivatives with respect to $x,y$. The derivations are
straightforward. In particular, it follows from Eqs.
(\ref{metrtensor}), (\ref{wip}) that the other derivatives of
$g_{00}$ are given by
\begin{equation}\begin{array}{c}
 g_{00,1,1}=\frac{2G}{c^2}\sum_k{\Biggl\{\frac{1}{\bigl[\bigl(x-x^{(k)}\bigr)^2+\bigl(y-y^{(k)}\bigr)^2
 +z^2\bigr]^{3/2}}}\\-\frac{3\bigl(x-x^{(k)}\bigr)^2}{\bigl[\bigl(x-x^{(k)}\bigr)^2+\bigl(y-y^{(k)}\bigr)^2
 +z^2\bigr]^{5/2}}\Biggr\}M^{(k)},\\
 g_{00,2,2}=\frac{2G}{c^2}\sum_i{\Biggl\{\frac{1}{\bigl[\bigl(x-x^{(k)}\bigr)^2+\bigl(y-y^{(k)}\bigr)^2
 +z^2\bigr]^{3/2}}}\\-\frac{3\bigl(y-y^{(k)}\bigr)^2}{\bigl[\bigl(x-x^{(k)}\bigr)^2+\bigl(y-y^{(k)}\bigr)^2
 +z^2\bigr]^{5/2}}\Biggr\}M^{(k)}.
\end{array}\label{wipc}\end{equation}

At the point $x=y=0$, the integration results in
\begin{equation}\begin{array}{c}
 g_{00,1,1}=
 g_{00,2,2}=\frac {2\pi G\mu R^2} {c^2(R^2+z^2)^{3/2}}.
\end{array}\label{wipk}\end{equation}

In the case under consideration,
\begin{equation}\begin{array}{c}
 R_{0110} =- R_{1010}=R_{0220} = -R_{2020}=\frac12g_{00,1,1}\\=\frac {\pi G\mu R^2} {c^2(R^2+z^2)^{3/2}}.
\end{array}\label{wiplm}\end{equation}

Other nonzero derivatives of
$g_{\mu\nu}$ calculated in
the same way are equal to
\begin{equation}\begin{array}{c}
 g_{11,1,1}=
 g_{22,2,2}=\frac {\pi G\mu R^2(R^2-8z^2)} {2c^2(R^2+z^2)^{5/2}},\\
g_{11,2,2}=
 g_{22,1,1}=\frac {3\pi G\mu R^4} {2c^2(R^2+z^2)^{5/2}},\\
g_{11,3,3}=
 g_{22,3,3}=-\frac {6\pi G\mu R^4} {c^2(R^2+z^2)^{5/2}},\\
g_{33,1,1}=
 g_{33,2,2}=\frac {6\pi G\mu z^2R^2} {c^2(R^2+z^2)^{5/2}},\\
g_{12,1,2}=-\frac {\pi G\mu R^2(R^2+4z^2)} {2c^2(R^2+z^2)^{5/2}},\\
g_{13,1,3}=g_{23,2,3}=-\frac {2\pi G\mu R^2(R^2-2z^2)} {c^2(R^2+z^2)^{5/2}}.
\end{array}\label{wipf}\end{equation}

These results allow us to calculate the other nonzero components of the Riemann tensor which are given by
\begin{equation}\begin{array}{c}
 R_{1221} =- R_{2121}=\frac {2\pi G\mu R^2} {c^2(R^2+z^2)^{3/2}},\\
R_{3113} =- R_{3131}=R_{2332} =- R_{3232}=-\frac {\pi G\mu R^2} {c^2(R^2+z^2)^{3/2}}.
\end{array}\label{wipln}\end{equation}

It is instructive to mention that $\pi\mu R^2=M$ where $M$ is the mass of the disk. Certainly, the Ricci tensor and the Ricci scalar curvature are equal to zero:
\begin{equation}\begin{array}{c}
 R_{\mu\nu} =0, \qquad \mathcal{R}=0.
\end{array}\label{Ricci}\end{equation}

Thus, the metric \eqref{wipgrav}, \eqref{wipnext} belongs to vacuum solutions of the Einstein equations. 
We underline an important contribution of the horizontal coordinates to the Riemann tensor. Taking into account this contribution substantially differs our approach from Refs. \cite{Taub,Taub_51,Vilenkin,EKasner} where the metrics depending on only one variable have been considered. We should mention that our approach based on the weak-field approximation cannot be applied on the plane $z=0$. On this plane, the surface energy-momentum tensor $S_{ik}$ should be used.

It is instructive to use the Riemann tensor for a calculation of second-order scalar invariants (on the disk axis). For the Kretschmann scalar, we get 
	\begin{equation}
	K_1=R_{\mu\nu\lambda\tau}R^{\mu\nu\lambda\tau}=\frac{48 G^2 \pi^2 R^4 \mu^2}{c^4(R^2+z^2)^3}.
	\label{Cinv}
	\end{equation}	
Comparing with the Schwarzschild solution,
		\begin{equation}
	K_1=\frac{48 G^2 M^2}{c^4r^6},
	\label{Cinv-2}
	\end{equation}
obviously reveals the absence of the singularity at $z=0$.
One more example for a comparison is the Taub solution \eqref{Taubm}, for which we obtain
\begin{equation}
K_1=\frac{3 K^4}{4(1+K |z|)^3}.
\label{Cinv-4}
\end{equation}
Notice the singularity at $|z|=-1/K,\, K<0$. As mentioned in Sec. \ref{reviews}, it does not belong to the coordinate singularity only.
Another invariant is the Chern-Pontryagin scalar and for the case of the disk we get
	\begin{equation}
	K_2=R_{\mu\nu\lambda\tau} {}^\star R^{\mu\nu\lambda\tau}=0,
	\label{Cinv-3}
	\end{equation}	
	where ${}^\star R^{\mu\nu\lambda\tau}$ is the left dual of the Riemann tensor.

Invariants \eqref{Cinv} -- \eqref{Cinv-3} preserve their structure up to coordinate transformation. Thus, it is possible to show that the disk field metric \eqref{wipgrav}, \eqref{wipnext} transformations to both the Schwarzschild and Taub solutions are inconsistent with the Kretschmann scalar invariant, demonstrating different field types.

The Chern-Pontryagin scalar \eqref{Cinv-3} relates to a gravitomagnetic field (see Sec. \ref{accelerationgravity}).

An analysis of Eqs. (\ref{Riemann30}), (\ref{wiplm}), and (\ref{wipln}) shows that the Riemann curvature describing effects
of the field inhomogeneity does not vanish near the disk surface, when
$z\rightarrow0$. The components of the Riemann tensor are of the order of $GM/(c^2R^3)$. We underline that the field inhomogeneity takes place even relative to the $x$ and $y$ directions. It vanishes only if the disk
radius tends to infinity ($R\rightarrow\infty$). It will be proven in
the next section that this situation is impossible.

It is instructive to compare the Riemann curvatures for the disk
and the Schwarzschild source. For this purpose, it is convenient
to use the standard definition of the gravitational radius
\begin{equation}\begin{array}{c}
r_g=\frac {2GM} {c^2}.
\end{array}\label{gradius}\end{equation}
For the disk,
\begin{equation}\begin{array}{c}
r_g=\frac {2\pi G\mu R^2} {c^2}.
\end{array}\label{raddisk}\end{equation} Thus, the order of magnitude of the nonzero components of the
Riemann tensor is $r_g/R^3$. For the Schwarzschild source, the
Riemann tensor components are of the order of $r_g/r^3$, where $r$ is
the distance from the source. We can note a substantial
difference between the two cases. The Riemann curvature decreases when $R$ (for the disk) and $r$ (for the Schwarzschild source) increase. In this case, the Newtonian acceleration remains
almost unchanged for the disk and decreases for the Schwarzschild source. This comparison shows that the gravitational fields of the
disk and the Schwarzschild source belong to different kinds of the
spatial inhomogeneity.

\section{Problem of existence of a uniform gravitational field}\label{existence}

It has been tacitly assumed in 
precedent investigations \cite{Kottler,Moller,Kottler2,Whittaker,Krige,Rohrlich,Jones,Munoz} that the infinite gravitating
plane can \emph{in principle} exist. However, an attentive look at
the problem brings the opposite conclusion.

The EEP compares inertial and gravitational phenomena when $\bm g$ is fixed and is equal to $-\bm a$. As follows from Eq. (\ref{Newtonian}), in this case $|\bm g|=2\pi G\mu$ near the disk surface. Thus, $\mu$ is to be fixed.
Equation
(\ref{raddisk}) shows that the ratio $r_g/R$ is proportional to
$R$. Even for the arbitrarily small surface mass density $\mu$, one can
find so large disk radius $R$ that $r_g/R>>1$. When $r>r_g$, a
field of a gravitating disk is equivalent to that of a pointlike
Schwarzschild source. The physical space created by this source
possesses the horizon at $r=r_g$. Therefore, all points near the
disk surface are \emph{under the horizon}. Since the physical
space near the disk is under the horizon even for finite (while
very large) $R$, it is senseless to consider the infinite
gravitating plane. This is a nonexistent object. Importantly, its nonexistence has been proven \emph{without the use of the weak-field approximation}. As a result, all
precedent investigations of the infinite gravitating plane
\cite{Kottler,Moller,Kottler2,Whittaker,Krige,Rohrlich,Jones,Munoz} have led to misleading results.


It is appropriate to note that, according to Eqs. \eqref{wipgrav} and
\eqref{wipnext}, the considered approximation ceases to be valid at
\begin{equation}
R\sim\frac{c^2}{2\pi G \mu}.
\end{equation}
At the same time, as follows from Eq. \eqref{raddisk}, this is exactly the condition for the disk to getting collapsed because  $R=r_g$. Thus, a criterion for the applicability of our approximation coincides with the condition of a gravitational stability of the disk.

For example, to estimate the discussed scales, the limiting radius of the disk is $R\sim 1 $ light year for surface density $\mu=10^9$  g/cm$^2$ and $\mu=10^3$ g/cm$^2$ for $R\sim 10^6 $ light years.

The problem of nonexistence of the infinite plane can be treated more rigorously if we consider  a plane with the energy density $\varrho$, pressure $p$, and look at the plane as an 2-brane embedded in the three-dimensional space \cite{mannheim2005brane}.
The energy-momentum tensor for the brane then takes a form:
\begin{equation}\label{tmn_brane}
T_{\mu\nu}^{brane}=\tau_{\mu\nu} \, \delta(z),
\end{equation}
where
\begin{equation}
\tau_{00}=\varrho,\qquad \tau_{11}=\tau_{22}=p.
\end{equation}

The most general metrics corresponding to the considered geometry is \cite{mannheim2005brane}
\begin{equation}\label{ds2_brane}
ds^2 = n^2(z)dt^2 - a^2(z)[dx^2+dy^2] - dz^2.
\end{equation}
The Einstein field equations with the metrics \eqref{ds2_brane} and the energy-momentum tensor \eqref{tmn_brane} lead to a single possible solution with exotic quintessence-like EoS \cite{mannheim2005brane,mannheim2000constraints}:
\begin{equation}\label{eos-brane}
p=-\frac{\varrho}{4}.
\end{equation}
We need to emphasize that this is the only static solution and all other solutions lead to unstable branes, which was proven by Vilenkin \cite{Vilenkin} for the particular case of the vacuum EoS $p=-\varrho$.
Dolgov and Khriplovich \cite{Dolgov} have come to this conclusion after the consideration of general properties of the field equation for the metric  \eqref{ds2_brane}.

Therefore, stabilization of the brane with a realistic EoS can be achieved only through additional matter sources in the bulk,  i.e. by the addition of the cosmological constant  $\Lambda$ into Eq. \eqref{tmn_brane}, as noted in Ref. \cite{Jones} (where the metrics with $n(z)=a(z)$ has been considered). 

The same problem arises in the brane-world models which consider embedding of the 4-dimensional Universe in the 5-dimensional anti-de Sitter bulk \cite{Randall:1,Randall:2}. In order to get a Poincare-invariant solution in these models, one needs to provide the brane with the tension $\sigma$ and set the cosmological constant $\Lambda$ in the 5-dimensional bulk, related to each other by some sort of fine tuning \cite{Rubakov:2001}:
\begin{equation}
\Lambda=-\frac{4\pi}{3}G_5\,\sigma^2.
\end{equation}
Here $G_5$ is the 5-dimensional gravitational constant.

Evidently, the arguments for the nonexistence of the infinite plane have a sufficiently general character and do not depend of any order of approximation.
Remarkably, an attempt to overcome the limitation  \eqref{eos-brane} leads to vanishing the curvature 
\cite{DaSilva}. In this case, the spacetime outside the plane is the Rindler spacetime \cite{DaSilva}. 

Thus, we should conclude that a gravitational field cannot be
uniform. In all probability, the quasi-uniform gravitational field
near the disk surface is the best approximation of the uniform
one. Nevertheless, the field inhomogeneity cannot be vanished and
the uniform gravitational field cannot be obtained with any real
gravitational sources. This means that the noncoincidence of spin
dynamics in an accelerated frame and a quasi-uniform (but
nevertheless nonuniform) gravitational field certified in Refs.
\cite{VarjuRyder,Lee} and confirmed in Refs.
\cite{MashhoonObukhov,PRDThomaspre} does not violate the EEP \cite{Einstein} formulated for the uniform
gravitational field.

\section{Particle motion and spin dynamics in the accelerated frame and in gravitational fields of several sources}\label{accelerationgravity}

The nonexistence of the uniform gravitational field does not
diminish the importance of the derivation of equations of particle
motion and spin dynamics in static gravitational fields and the
comparison of these equations with corresponding ones for the
accelerated frame. One can use the well-known general equations of particle
motion (\ref{Lorcova}) and (\ref{LorChrc}).
Evidently, the equations of particle motion in the accelerated
frame and in any static gravitational field must differ. Their
equivalence could take place only in the case of the coincidence
of \emph{all} components of the metric tensor. Otherwise, spatial
components of the metric tensor vary for different gravitational
fields even if the Newtonian limit of these fields is the same.
Therefore, corresponding equations of particle motion do not
coincide either.

It has been demonstrated in Ref. \cite{PRDThomaspre} that the
difference takes place even for the Schwarzschild field in the
Cartesian and isotropic coordinates. In the weak-field
approximation, the metric tensors of the Schwarzschild field can
be given by \cite{LL2}
\begin{equation}
g^{(C)}_{00}=1-\frac{r_g}{r},\quad g^{(C)}_{0i}=0,\quad
g^{(C)}_{ij}=-\left(\delta_{ij}-\frac{r_gx_ix_j}{r^3}\right)
\label{mtCar}\end{equation} and
\begin{equation}
g^{(I)}_{00}=1-\frac{r_g}{r},\quad g^{(I)}_{0i}=0,\quad
g^{(I)}_{ij}=-\left(1-\frac{r_g}{r}\right)\delta_{ij}
\label{mtiso}\end{equation} 
in the Cartesian and isotropic
coordinates, respectively. Here $r_g=2GM/c^2$ is the gravitational
radius. The equations of the particle
motion in the Cartesian coordinates read \cite{PRDThomaspre}
\begin{equation}
\begin{array}{c}
\frac{du_{i}}{ds}=\frac{(u^0)^2r_g}{2r^3}\left\{x^i\left[1+\frac{3(\bm\beta\cdot\bm r)^2}{r^2}\right]-2\beta^i(\bm\beta\cdot\bm r)\right\}\\=-\frac{(u^0)^2}{c^2}\left\{g^i\left[1+\frac{3(\bm\beta\cdot\bm r)^2}{r^2}\right]-2\beta^i(\bm\beta\cdot\bm g)\right\},\qquad \frac{du_{0}}{ds}=0,
\end{array} \label{LorcCar} \end{equation}
\begin{equation}
\begin{array}{c}
\frac{du^{i}}{ds}=-\frac{(u^0)^2r_g}{2r^3}x^i\left[1+2\bm\beta^2-\frac{3(\bm\beta\cdot\bm r)^2}{r^2}\right]=\frac{(u^0)^2}{c^2}g^i\left[1+2\bm\beta^2-\frac{3(\bm\beta\cdot\bm r)^2}{r^2}\right],\\ \frac{du^{0}}{ds}=-\frac{(u^0)^2r_g(\bm\beta\cdot\bm r)}{r^3}=2\frac{(u^0)^2}{c^2}(\bm\beta\cdot\bm g).
\end{array} \label{LorcCap} \end{equation}
We do not make a difference between the upper and lower indices for the Newtonian acceleration $\bm g$.

The corresponding equations in the isotropic coordinates have the form \cite{PRDThomaspre}
\begin{equation}
\begin{array}{c}
\frac{du_{i}}{ds}=\frac{(u^0)^2r_g}{2r^3}x^i\left(1+\bm\beta^2\right)=-\frac{(u^0)^2}{c^2}g^i\left(1+\bm\beta^2\right),\qquad \frac{du_{0}}{ds}=0,
\end{array} \label{Lorciso} \end{equation}
\begin{equation}
\begin{array}{c}
\frac{du^{i}}{ds}=-\frac{(u^0)^2r_g}{2r^3}\left[x^i(1+\bm\beta^2)-2\beta^i(\bm\beta\cdot\bm r)\right]=\frac{(u^0)^2}{c^2}\left[g^i(1+\bm\beta^2)-2\beta^i(\bm\beta\cdot\bm g)\right],\\ \frac{du^{0}}{ds}=-\frac{(u^0)^2r_g(\bm\beta\cdot\bm r)}{r^3}=2\frac{(u^0)^2}{c^2}(\bm\beta\cdot\bm g).
\end{array} \label{Lorcisp} \end{equation}

These equations can be compared with the related equations for the uniformly accelerated frame,
\begin{equation}
\begin{array}{c}
\frac{du_{i}}{ds}=\frac{(u^0)^2a^i}{c^2},\qquad \frac{du_{0}}{ds}=0,
\end{array} \label{Loraiso} \end{equation}
\begin{equation}
\begin{array}{c}
\frac{du^{i}}{ds}=-\frac{(u^0)^2a^i}{c^2},\qquad \frac{du^{0}}{ds}=-2\frac{(u^0)^2}{c^2}(\bm\beta\cdot\bm a).
\end{array} \label{Loraisp} \end{equation}

The comparison of Eqs. (\ref{LorcCar}) -- (\ref{Loraisp}) shows
that the terms different for the Schwarzschild field and the
uniformly accelerated frame ($\bm g=-\bm a$) are of the same order
of magnitude as the corresponding terms different for the
Schwarzschild field in the Cartesian and isotropic coordinates. It
has been concluded in Ref. \cite{PRDThomaspre} that the spatial
inhomogeneity significantly influences the form of the equations
of motion and the results presented above do not give a reason for
the assertion about a violation of the EEP.

The consideration of the quasi-uniform gravitational field of the
disk confirms this conclusion. In this case, the equations of
motion take the form
\begin{equation}
\begin{array}{c}
\frac{du_{0}}{ds}=
0,\qquad
\frac{du_{1}}{ds} = g_{31,1}u^1u^3,\qquad
\frac{du_{2}}{ds}=g_{32,2}u^2u^3,
\\
\frac{du_{3}}{ds}=\frac12\left[g_{00,3}(u^0)^2+g_{11,3}(u^1)^2+g_{22,3}(u^2)^2+g_{33,3}(u^3)^2\right],\qquad
\end{array} \label{qualo} \end{equation}
\begin{equation}
\begin{array}{c}
\frac{du^{0}}{ds}=-g_{00,3}u^0u^3,\qquad
\frac{du^{1}}{ds}=g_{11,3}u^1u^3,\qquad
\frac{du^{2}}{ds}=g_{22,3}u^2u^3,\\
\frac{du^{3}}{ds}=\frac12\left[-g_{00,3}(u^0)^2-
(g_{11,3} -2 g_{31,1})(u^1)^2-
(g_{22,3} -2 g_{32,2})(u^2)^2+
g_{33,3}(u^3)^2\right],\qquad
\end{array} \label{quaup} \end{equation} where the derivatives of the
metric tensor components are given by Eq. (\ref{wipnder}).

In the case of $z=0$, these equations reduce to
\begin{equation}
\begin{array}{c}
\frac{du_{0}}{ds}=\frac{du_{1}}{ds}=\frac{du_{2}}{ds}=0,\qquad
\frac{du_{3}}{ds}=\frac{(u^0)^2}{c^2}\left[g_{\pm}(1+\bm\beta^2)
\right],
\end{array} \label{Lorclow} \end{equation}
\begin{equation}
\begin{array}{c}
\frac{du^{1}}{ds}=-2\frac{(u^0)^2}{c^2}\beta^1(\bm\beta\cdot\bm g),\qquad
\frac{du^{2}}{ds}=-2\frac{(u^0)^2}{c^2}\beta^2(\bm\beta\cdot\bm g),\qquad
\frac{du^{3}}{ds}=-\frac{(u^0)^2}{c^2}g_{\pm}(1+\bm\beta^2),\\
\frac{du^{0}}{ds}=2\frac{(u^0)^2}{c^2}(\bm\beta\cdot\bm g),
\end{array} \label{Lorcupp} \end{equation}
where $g_{\pm}=\pm2\pi G\mu$ correspond to limits in Eq. (\ref{Newtonian}) at $z\rightarrow \pm 0$.

We should not confuse $\beta^i=u^i/u^0$ and the square of $\bm\beta$.

In the case under consideration, the Newtonian acceleration defined by Eq. (\ref{Newtonian}) has two values describing the acceleration on each side of the disk at $z=0$:
\begin{equation}\begin{array}{c} \bm g= - g_{\pm}\,\bm e_z. \qquad
\end{array}\label{Newtonatn}\end{equation}

We can conclude that the particle motion in the quasi-uniform gravitational field of the disk and in the cases listed above substantially differs.
In particular, the horizontal forces, $m\frac{du^{1}}{ds}$ and $m\frac{du^{2}}{ds}$, acting on the particle in the gravitational field of the disk are absent in the accelerated frame at $\bm a=-\bm g$ and in the Schwarzschild field in the Cartesian coordinates. In the Schwarzschild field in the isotropic coordinates, the horizontal forces are the same as in the field of the disk at $z=0$ but the equations for $du^{3}/(ds)$ and $du_{3}/(ds)$ substantially differ.

Let us also consider the spin motion. It can be properly described with the use of the gravitoelectromagnetic fields first introduced by Pomeransky and Khriplovich \cite{PK} (see Ref. \cite{PRDThomaspre} and references therein). In the general form, the angular velocity of the spin rotation is given by
\begin{equation}
\begin{array}{c} \frac{d\bm\zeta}{d\tau}=\bm{\Omega}\times \bm\zeta, ~~~ \bm{\Omega}=-\bm{\mathcal{B}}+
\frac{\widehat{\bm u}\times\bm{\mathcal{E}}}
{u^{\widehat{0}}+1},
\end{array} \label{omgemtau}
\end{equation} where $\bm\zeta$ is the spin (pseudo)vector, $\tau$ is the proper time and $\bm{\mathcal{E}}$ and $\bm{\mathcal{B}}$ are the gravitoelectric and gravitomagnetic fields. In the weak-field approximation, these fields are defined by \cite{Warszawa,PRDThomaspre}
\begin{equation}
\begin{array}{c} \mathcal{E}_{\widehat{i}}=-\frac{c}{2}\left[g_{00,i}u^{\widehat{0}}+
\left(g_{0i,j}+
g_{0j,i}-g_{ij,0}\right)u^{\widehat{j}}\right],\\
\mathcal{B}_{\widehat{i}}=\frac{c}{4}e_{ijk}\left[(g_{0j,k}-g_{0k,j})
u^{\widehat{0}}+\left(g_{jl,k}-g_{kl,j}\right)u^{\widehat{l}}\right].
\end{array} \label{explg}
\end{equation}

For the Schwarzschild field in the isotropic coordinates, the gravitoelectric and gravitomagnetic fields are equal to \cite{PRDThomaspre}
\begin{equation}\begin{array}{c}
\bm{\mathcal{E}}=-\frac{GM\bm r}{cr^3}u^{\widehat{0}}=\frac{\bm gu^{\widehat{0}}}{c}, ~~~ \bm{\mathcal{B}}=-\frac{GM}{cr^3}\bm r\times\widehat{\bm u}=\frac{\bm g\times\widehat{\bm u}}{c},
\end{array}\label{gemfSch}\end{equation} where $\bm g$ is the Newtonian acceleration. For the Schwarzschild field in the Cartesian coordinates,
the gravitoelectromagnetic fields are the same \cite{PRDThomaspre}. Thus, the
equations of the spin motion in the Schwarzschild field
have the same form in the Cartesian and isotropic coordinates.

In the same approximation, the gravitoelectromagnetic fields in the uniformly accelerated
frame are given by
\begin{equation}\begin{array}{c}
\bm{\mathcal{E}}=-\frac{\bm au^{\widehat{0}}}{c}, ~~~ \bm{\mathcal{B}}=0.
\end{array}\label{gefua}\end{equation}

Equations (\ref{gemfSch}) and (\ref{gefua}) show the significant difference between the gravitoelectromagnetic fields in the uniformly accelerated frame and in the Schwarzschild spacetime. The gravitomagnetic field in the Schwarzschild spacetime, contrary to the uniformly accelerated frame, is nonzero.

The calculation of the gravitoelectromagnetic fields for the quasi-uniform gravitational field of the disk brings the following result:
\begin{equation}\begin{array}{c}
\mathcal{E}_{\widehat{1}}=\mathcal{E}_{\widehat{2}}=0,\qquad \mathcal{E}_{\widehat{3}}=-\frac c2g_{00,3}u^0,\\
\mathcal{B}_{\widehat{1}}=\frac c2 (g_{22,3}-g_{32,2}) u^2=\frac c2 g_{00,3} u^2,~
\mathcal{B}_{\widehat{2}}=-\frac c2 (g_{11,3}-g_{31,1}) u^1=-\frac c2 g_{00,3} u^1,~
\mathcal{B}_{\widehat{3}}=0,
\end{array}\label{gefqu}\end{equation}
where the metrics symmetry
\eqref{gnozero} and \eqref{gmnDerivDepend} is used.

Amazingly, in the case of $z=0$ we obtain the equation similar to Eq. (\ref{gemfSch})
\begin{equation}\begin{array}{c}
\bm{\mathcal{E}}=\frac{\bm gu^{\widehat{0}}}{c}, ~~~ \bm{\mathcal{B}}=\frac{\bm g\times\widehat{\bm u}}{c}.
\end{array}\label{gemfScn}\end{equation} Here $\bm g$ is defined by Eq. \eqref{Newtonatn}. The
dependence of the gravitoelectromagnetic fields on the $z$-reflection, $\bm{\mathcal{E}}\rightarrow -\bm{\mathcal{E}}$ and  $\bm{\mathcal{B}}\rightarrow -\bm{\mathcal{B}}$ when $z\rightarrow-z$, is taken into account.

Now we can relate the gravitomagnetic field $\bm{\mathcal{B}}$ from \eqref{gemfScn} with the second order scalars \eqref{Cinv} and \eqref{Cinv-3}. Since $K_1>0$ and $K_2=0$, we can note the existence of an observer for which the gravitomagnetic field disappears \cite{gravi_electro_mag}. We see that it is indeed the case for the coaxial moving, when $\bm g \parallel\widehat{\bm u}$.

Thus, the gravitoelectromagnetic fields in the Schwarzschild spacetime in the Cartesian and isotropic coordinates
and in the quasi-uniform gravitational field of the disk are given by the same expressions in terms of the Newtonian acceleration. As a contrary, the gravitoelectromagnetic fields in the accelerated frame are different (the gravitomagnetic field is equal to zero).

\section{Discussion and summary}\label{discussion}

The metrics (\ref{Kottler}) and (\ref{Kottler2}) obtained in Refs. \cite{Kottler,Moller,Kottler2,Whittaker,Rohrlich,Krige} are inappropriate for a description of a uniform gravitational field because they describe flat spacetimes and do not reproduce the correct weak-field approximation defined by Eqs. (\ref{wipgrav}), (\ref{wipnext}). The careful analysis of this problem fulfilled in our work needs taking into account a dependence on three spatial coordinates [see, e.g., Eqs. (\ref{wip}) and (\ref{gnozero0})], while the metrics considered in Refs. \cite{Taub,Taub_51,Vilenkin,EKasner} depend only on one variable.

There is no problem of gravitation singularity in the weak-field approximation, because the criterion of its applicability coincides with the condition of a gravitation collapse of the disk.

The Kottler-Whittaker metric (\ref{Kottler2}) can be considered as a limiting case of the Schwarzschild spacetime \cite{Karlov}. However, the real metric of a large disk differs from the metrics (\ref{Kottler}) and (\ref{Kottler2}). In particular, all four diagonal components of the metric (\ref{wipgnear}) are nontrivial. A unbounded increase of the disk radius brings the metric components $g_{00},~g_{11}$, and $g_{22}$ to the infinity.

The derivatives of the metric components defined by Eq. (\ref{wipnder}) also disagree with previously obtained results \cite{Kottler,Moller,Kottler2,Whittaker,Rohrlich,Krige}. As opposed to our results, only the derivatives of two diagonal components of the metric (\ref{Kottler2}) are nontrivial and the related derivatives of spatial components do not coincide (they have different signs). Just the derivatives of the metrics define equations of motion.

There were a lot of attempts to associate the Taub solution \eqref{Taubm} with different kinds of realistic field sources. However, the Taub solution belongs to the general form of the plane-symmetric solution \eqref{ds2_brane}, and the corresponding stable plane can be only exotic with the quintessence-like EoS \eqref{eos-brane}: $p=-\rho/4$. 
Moreover, the weak-field limit of the Taub metric can be generated by a plane with the same EoS, as shown by Vilenkin \cite{Vilenkin}.
At the same time, the Taub solution is singular.
These circumstances prevent to construct physically acceptable plane as the source of the Taub metric. On the other hand, one can consider a plane with a positive mass and a positive pressure. However, it is still unclear (due to curvature vanishing \cite{DaSilva}) how to combine a static character of the solution and to avoid the gravitational collapse of the infinite plate.  
The only possibility to partially exploit the Taub metric in the context of realistic sources is the use of an extended solution with the general metric \eqref{ds2_brane} for а slice with a \textit{nonzero} thickness. Resulting solutions lead to interesting configurations with two kinds of boundary conditions. They are associated with the Rindler metric and the Taub one, respectively \cite{Novotny,grav_slab,reint_taub,Burcev,Arutyunyan,Avakyan,Homogeneous_grav_field}.
A geometry of these configurations consists in the flat space with the Rindler metric from one side of the slice and with the Taub metric from another side. An unexpected peculiarity of the solutions is loosing of mirror symmetry \cite{Novotny,grav_slab}. A slice with a matter is inside a singular surface. 
Contrary to this, the solution \eqref{wipgrav}, \eqref{wipnext} is regular, naturally preserves mirror symmetry, satisfies all the energy conditions and opens a way to find a full general-relativistic solution for the disk, at least, numerically (see Appendix).
We can note that the disk field and the field with the Taub metric are qualitatively different as follows from the second-order invariants.

We can certify that the spacetime attributed to the large disk is essentially curved. At some time, an increase of the disk radius leaves all the disk under the horizon. Therefore, the disk radius is limited. As a result, the Riemann tensor and the field inhomogeneity cannot be vanished.
The Ricci tensor is zero. The existence of a horizon at a large disk radius is the main disagreement with the previous results \cite{Kottler,Moller,Kottler2,Whittaker,Krige,Rohrlich,Jones,Munoz}. Even for the arbitrarily small surface mass density, one can find so large disk radius $R$ that $r_g/R>>1$. In this case, the
field of the gravitating disk is equivalent to that of a pointlike Schwarzschild source and all points near the disk surface are \emph{under the horizon}. Therefore, the infinite gravitating plane is a nonexistent object. Importantly, the nonexistence of the uniform gravitational field has been proven \emph{without the use of the weak-field approximation}. The same result follows from the brane-world point of view, where the infinite plane is embedded in 3D space.

This conclusion is very important for the correct analysis of the EEP. Sec. \ref{accelerationgravity} presents a comparison of the equations of particle motion in the three important quasi-uniform gravitational fields and in the uniformly accelerated frame. We have considered the Schwarzschild field in the Cartesian and isotropic coordinates
and the field of the gravitating disk. The equations of particle motion in all four cases do not
coincide with each other while the Newtonian limit in these cases is the same. Therefore, the spatial inhomogeneity
significantly influences the form of the equations of motion. Since the EEP has been formulated only relative to a constant \emph{uniform} gravitational field \cite{Einstein}, it is not violated by the results presented. Thus, the consideration of the quasi-uniform gravitational field of the disk confirms the conclusion first made in Ref. \cite{PRDThomaspre}.

It is also important that the gravitoelectromagnetic fields in the Schwarzschild spacetime in the Cartesian 
and isotropic coordinates and in the quasi-uniform gravitational field of the disk are defined
by the same expressions in terms of the Newtonian acceleration. Therefore, the corresponding equations of the spin motion coincide. The equation of the spin motion in the accelerated frame significantly differs because the gravitomagnetic field is equal to zero.

We should underline that the results obtained in the present paper cover only the weak-field approximation while the general case needs a separate analysis.
Summarizing these results, we can note that the investigation of the quasi-uniform gravitational field of the disk is rather important for an analysis of fundamental problems of contemporary gravity.
In particular, our results show that the EEP indicates the equivalence of \emph{a physical nature} of inertia and gravity rather that the equivalence of observable effects.

\section*{Acknowledgments}

The authors are grateful to F. W. Hehl and Yu. N. Obukhov for useful discussions. The work was supported in part by the Belarusian Republican Foundation for Fundamental Research
(Grant No. $\Phi$18D-002), by the National Natural Science
Foundation of China (Grant No. 11575254),
by the National Key Research and Development Program of China
(No. 2016YFE0130800),
and by the Heisenberg-Landau program of the German Federal Ministry
of Education and Research (Bundesministerium f\"{u}r Bildung und
Forschung). A. J. S. also acknowledges hospitality and support by the
Institute of Modern
Physics of the Chinese Academy of Sciences.


\section*{Appendix: relation to the canonical Weyl metric}
\label{appx}
%
Let us show how the disk solution \eqref{wipgrav}, \eqref{wipnext} is related to the canonical Weyl metric for the axially symmetric field \eqref{dsaxial_0}. 
%
%

First of all, we can expand Weyl's metric function $e^{2\psi}$ from \eqref{dsaxial_0} in the weak-field approximation and can write in the leading order \cite{griffiths2009exact}:
\begin{equation}\label{ap-ds_lin}
ds^2 = (1+2\psi)dt^2 - (1-2\psi)\left[e^{2\sigma}(d\rho^2+dz^2)+\rho^2 d\varphi^2\right].
\end{equation}
In this form, some similarity to the weak-field metric in isotropic spatial coordinates should be noted.
Indeed, we have $\sigma \xrightarrow[\rho \rightarrow 0]{} 0$ according to the properties of the Weyl solution. Thus, we get $e^{2\sigma}=1$ on the axis and the metric $\eqref{ap-ds_lin}$ takes exactly isotropic form in quasicylindrical coordinates: 
\begin{equation}\label{ap-ds_lin2}
ds^2 \left|_{\rho=0}\right. = (1+2\psi)dt^2 - (1-2\psi)\left[d\rho^2 + dz^2+\rho^2 d\varphi^2\right].
\end{equation}

According to Eq. \eqref{ap-ds_lin2}, we obtain the important conclusion: 
\textit{
in the weak-field approximation, the Weyl solution \eqref{dsaxial_0} for an axially symmetric gravitational field reduces to the conformal Euclidean metric (on the axis) and the canonical quasicylindrical coordinates $\{\rho, z, \varphi\}$ are transformed to a usual isotropic cylindrical  coordinates in a flat space.}
Next point is that $\psi(\rho,z)$ reduces to the Newtonian gravitational potential of the source of axially symmetric field.

This property of the Weyl metric can be clearly illustrated with the Schwarzschild solution for the mass $M$. With the use of Ref. \cite{griffiths2009exact}, it can be expressed in the Weyl form as follows:
\begin{equation}\label{ap-Schw}
\psi_S=\frac{1}{2}\ln\frac{R_+ + R_- -2 M}{R_+ + R_- +2 M}, \,~~ 
\sigma_S=\frac{1}{2}\ln \frac{(R_+ + R_-)^2 -4 M^2}{4R_+ R_-},~~
\end{equation}
where
\begin{equation}
R_{\pm}=\sqrt {\rho ^{2}+(z\pm M)^{2}}.
\end{equation}
In the weak-field approximation at small $M$, one has
\begin{equation}\label{ap-Schw2}
\begin{array}{c}
\psi_{S}=-\frac{M}{\sqrt{\rho^2+z^2}}+O(M^2), \\
\sigma_{S}=-\frac{\rho^2 M^2}{2(\rho^2+z^2)^2}+O(M^3).
\end{array}
\end{equation}
As expected, one recognizes in $\psi_{S}$ the usual Newtonian potential of the central point and $\sigma_{S}=0$  when $\rho =0$.
A source rod spanning $z\in[-M,M]$ in the complete Weyl solution (see the beginning of Sec. \ref{disk_field}) reduces now to a point source in the limit $M \ll 1$.
Finally, one can see that the canonical coordinates became the usual cylindrical coordinates in the conformal flat space and 
the functions $\psi_{S}$ and $\sigma_{S}$ are indeed solutions of the vacuum field equations \eqref{dsaxial_1}, \eqref{dsaxial_2}
for the Weyl metric\footnote{
Another equation, $\sigma_{,\rho\rho}+\sigma_{,zz}+(\nabla\psi)^2=0$, is a direct consequence of the equations considered.
}.

It is now clear how to write a uniform disk metric in the weak-field approximation. We can turn back to the Schwarzshild mertic \eqref{wip}
in Cartesian coordinates and can transform it to isotropic one via \cite{LL2}
\begin{equation}\label{ap-trans}
h_{\mu\nu}^{'}=h_{\mu\nu}-\frac{\partial\xi_\mu}{\partial x^\nu}-\frac{\partial \xi_\nu}{\partial x^\mu},
\end{equation}	
where 
\begin{equation}\label{ap-trans2}
\xi^k = -\frac{r_g x^k}{2 r},
~~k=\{1,2,3\}
\end{equation}	
and $r_g$ is the gravitational radius \eqref{gradius}.
For the point source, we have therefore the required diagonal form
of the metric:
\begin{equation}\label{ap-ds_lin3}
ds^2 
= (1+2\psi)dt^2 - (1-2\psi)\left[dx'\,^2 + dy'\,^2 + dz'\,^2\right],
\end{equation}
where the new and old coordinates are related via $x'\, ^\mu = x^\mu + \xi^\mu $ and $\psi $ is the Newtonian potential of a point source which is equal to $\phi^{(i)}$ in Eq. \eqref{wip}.

Recall that the metric of the disk field in pure Cartesian coordinates \eqref{wipgrav}, \eqref{wipnext} has been obtained by integrating the field \eqref{wip} of the point source. 
In the same way, the disk field in isotropic Cartesian coordinates follows from the integration of the metric \eqref{ap-ds_lin3}. We can see that $h'_{00}=h_{00}$ and $\psi=\psi(z)=\phi(z)$ from Eq. \eqref{wipgrav} is the Newtonian potential on the axis of the disk.
Introducing cylindrical  coordinates $\{\rho, z, \varphi\}$ instead of isotropic Cartesian ones, one gets to the canonical Weyl metric in the weak field defined by Eq. \eqref{ap-ds_lin2}. 

\begin{figure}[t]
	\centering
	\includegraphics[width=12 cm]{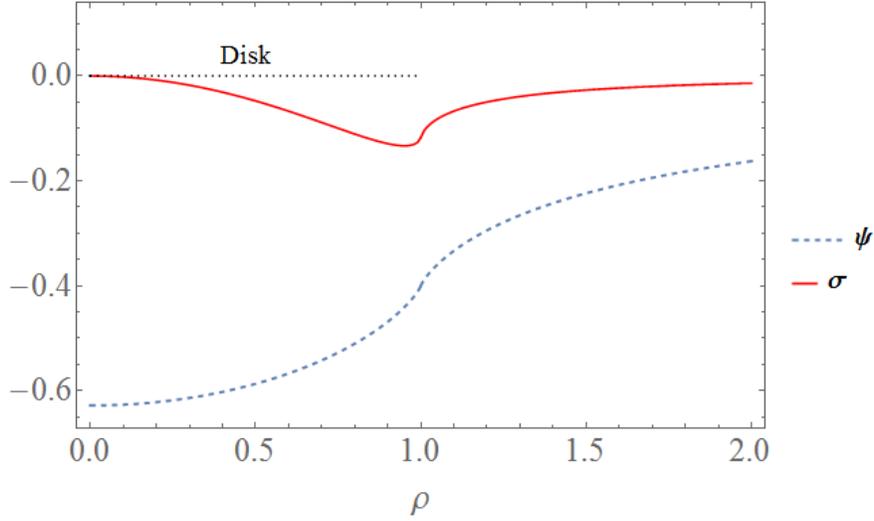}
	\caption{The radial profile of the functions $\psi(\rho,z_0)$ and $\sigma(\rho,z_0)$ for the disk in the weak-field approximation. The disk radius is $R=1$ and $z_0=10^{-4}$.}
	\label{fig1}
\end{figure}

\begin{figure}[h]
	\centering
	\includegraphics[width=12 cm]{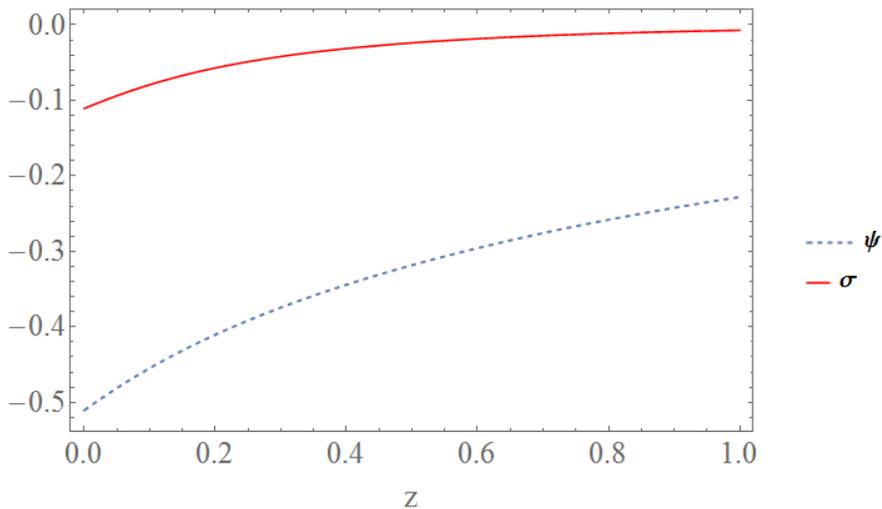}
	\caption{The profile of the functions $\psi(\rho_0,z)$ and $\sigma(\rho_0,z)$ along the disk axis in the weak-field approximation. The disk radius is $R=1$ and $\rho_0=0.8$.}
	\label{fig2}
\end{figure}

We can now write a full solution for the disk including the off-axis space.
It is mentioned in Sec. \ref{disk_field} that  $\phi(z)$ from Eq. \eqref{wipgrav} is the on-axis limit of the full Newtonian potential $\psi_N$ of the disk in the whole space. If we take $\psi(\rho,z)=\psi_N$, we can calculate second canonical Weyl function $\sigma(\rho,z)$ using the field equations [see Eq. \eqref{dsaxial_2}]:
\begin{equation}\label{ap-sigma}
\sigma(\rho,z) = \int\limits_{\Gamma} \rho\left[ (\psi_{,\rho}^2-\psi_{,z}^2)d\rho+2\psi_{,\rho}\psi_{,z}dz\right],
\end{equation}
where the path $\Gamma$ covers the disk. Details of the integration can be found in Ref. \cite{Synge}.
The full disk potential $\psi_N$ is quite cumbersome (see, for instance, Ref. \cite{lass1983gravitational}) and includes elliptic integrals. Therefore, it is better to solve Eq. \eqref{ap-sigma} numerically. The figure \ref{fig1} illustrates a radial behaviour of the Weyl potentials $\psi, \sigma$ and the figure \ref{fig2} shows the profiles along the disk axis ($G=1, c=1$). 

In this way, we obtain the full Weyl solution for the disk in the canonical form \eqref{ap-ds_lin} satisfying the vacuum field equations \eqref{dsaxial_1}, \eqref{dsaxial_2}.

It is clear from the weak-field analysis that the full general relativistic solution $\psi_{GR}$ can be constructed from the Newtonian potential $\psi_N$ as the limit of the exponential expansion
\begin{equation}
\label{psiGR}
\exp(2\psi_{GR})=\sum_{n=0}^{\infty}\frac{(2\psi_N)^n}{n!}.
\end{equation}	
This is exactly an infinite series of the uniform disk solution which has been mentioned in Ref. \cite{Morgn}. A functional representation of the limit \eqref{psiGR} is still an open question.




						
						
						
				






\end{document}